\documentclass[aps,showpacs, nofootinbib]{revtex4}

\newcommand{\be}{\begin{equation}}
\newcommand{\ee}{\end{equation}}
\newcommand{\ben}{\begin{eqnarray}}
\newcommand{\een}{\end{eqnarray}}

\newcommand{\cA}{{\cal A}}
\newcommand{\cT}{{\cal T}}

\newcommand{\cL}{{\cal L}}
\newcommand{\cH}{{\cal H}}

\newcommand{\p}{\partial}
\newcommand{\na}{\nabla}

\newcommand{\ep}{\epsilon}
\newcommand{\bep}{\bar \epsilon}

\newcommand{\bdel}{\bar \delta}

\pacs{04.50.+h}

\begin{document}

\title{First Law of p-brane Thermodynamics}

\author{Marek Rogatko}
\affiliation{Institute of Physics \protect \\
Maria Curie-Sklodowska University \protect \\
20-031 Lublin, pl.~Marii Curie-Sklodowskiej 1, Poland \protect \\
rogat@kft.umcs.lublin.pl \protect \\
marek.rogatko@poczta.umcs.lublin.pl}
\date{\today}

\begin{abstract}
We study the {\it physical process} version and {\it equilibrium state} version of the first law
of thermodynamics for a charged p-brane.
The general setting for our investigations is $(n + p + 1)$-dimensional Einstein dilaton gravity
with $(p + 2)$ strength form fields.
 \end{abstract}

\maketitle

\section{Introduction}
The unrelenting search for the unification scheme in contemporary high-energy physics leads
to the idea that our Universe may have been more than four-dimensional.
One of the most promising approaches to the unification of fundamental forces of nature
is the superstring/ M-theory, which is formulated on a higher dimensional manifold. On the
other hand, black holes through their quantum behaviour and thermodynamical properties
have played important role in our quest for understanding quantum theory of gravity. It
triggers continuously growing interests in studying properties of black holes and other black objects
appearing in higher dimensional theories of gravity.
\par
The uniqueness theorem for higher dimensional static black holes is quite well justified \cite{uniq}.
On the contrary, the situation is far from obvious as far as stationary axisymmetric
higher dimensional black holes is concerned. It was shown that even in five-dimensions
a kind of black object appeared. It was called {\it black ring} and its topology
of the event horizon is $S^2 \times S^1$ \cite{emp02}. This object is equipped with the same
mass and angular momentum as five-dimensional spherically symmetric stationary axisymmetric black
hole. For a {\it black rings} story see Ref.\cite{emp06} and references therein. However,
the assumption about topology of the considered black object enables one
to prove uniqueness theorem for five-dimensional vacuum stationary axisymmetric black hole \cite{mor04}
and for stationary axisymmetric self-gravitating $\sigma$-models \cite{rog04a}. Taking into account
the so-called {\it rod structure} \cite{har} enables to broaden these attempts to the case of 
asymptotically flat five-dimensional black hole solutions
of vacuum Einstein Eqs. \cite{hol07a} and charged
five-dimensional stationary axisymmetric black hole \cite{hol07b} (where the gauge field appeared only in fifth dimension).
In Ref.\cite{mor08} that assuming the existence of two additional 
commutating axial Killing vector fields and the horizon topology of black ring $S^1 \times S^2$,
the only asymptotically flat black ring solution with a regular horizon is the Pomeransky-Sen'kov (PS)
black ring \cite{pom06}. In Ref.\cite{rog08} it was show that 
in five-dimensional admitting the self-gravitating $\sigma$-model                         
the only asymptotically flat black
ring with a regular rotating event horizon is the black ring characterized by mass and two angular
momenta with constant mapping.
\par
However, in multi-dimensional theories 
the situation drastically changes. One can consider product of $(d-m)$-dimensional Minkowski spacetime 
times a compact Ricci flat $m$-dimensional manifold $Ricci^{m}$.
It happens that black objects with different horizon topology 
depending on the size of extra dimensions arise. Namely, when the size of a compact manifold is large comparing
to the event horizon of black object, one obtains a black hole with topology of the event horizon $S^{d-2}$.
On the contrary, when the size of the manifold is small one gets a black string with $S^{d-m-2} \times Ricci^{m}$.
Such kinds of black object were intensively studied in five-dimensions ($Mink^{4} \times S^{1}$) \cite{five}.
In the spacetime in question, the arising black holes were named {\it caged black holes}. Their numerical
studies were presented in Refs.\cite{num}, while Ref.\cite{anal} was devoted to their analytical studies.
\par
A string solution with topology $S^{d-3} \times S^1$ is $z$-dependent and it is described by $(d-1)$-dimensional
Schwarzschild solution with coordinate $dz^2$. 
It turns out that non-extremal stationary {\it translationally invariant} branes
are unstable. In Ref.\cite{gre93}
the authors investigated the stability of a black p-brane and revealed that such
a background was unstable as the compactification scale of extended directions
became larger than the order of the horizon radius. Next, it was shown in \cite{gre94} for a class
of a magnetically charged p-brane solutions of stringy action that the instability
persisted to appear but decreased with the charge increase to the extended value.
On the other hand, it happened that branes with extremal charge were stable \cite{gre95}.
It was conjectured \cite{gub00} that for a black brane with a translationally
symmetry Gregory-Laflamme (GL) instabilities occurred when the brane in question
was thermodynamically unstable. It was also demonstrated numerically that a certain
class of black holes in AdS spacetime were unstable against linear perturbations \cite{gub01}. 
Then, in Ref.\cite{rea01} a strong support for this conjecture was given. Among all,
it was shown that it was possible to reveal that many black branes were classically
unstable without implementing arduous numerical analysis of the problem.
\par
In Ref.\cite{hor01} it was revealed that non-extremal {\it translationally invariant}
black brane decayed to some stable stationary configuration. It settled down
to a stable but inhomogeneous p-brane. It was also found \cite{hor02}
that there existed a large class of new stable inhomogeneous black brane solutions
which were unrelated to the GL instability. They exist even if the adequate
homogeneous solution with the same mass and charge is stable.
\par
But in principle it could be possible to derive the first law of thermodynamics
for branes that are not {\it translationally invariant}. Some progress in this direction
was made. In Ref.\cite{tra04} a black brane spacetime that had at least one spatial
translation Killing field tangent to the brane was considered. For static
charge black brane a law which related the tension
perturbation, surface gravity, charge of the event horizon area and
variations of charges and their currents was derived. On the other hand,
in Ref.\cite{har04} a generalization of the gravitational tension in a given
asymptotically {\it translationally invariant} spacetime
direction was presented. The tension was defined in analogy with
the Hawking-Horowitz energy definition \cite{haw96}. This definition was applied
for finding a general tension formula in the case for near-extremal branes.
\par
Higher dimensional gravity 
possess also more complicated black objects such as
p-branes, black strings, black Saturn, i.e., $n$-dimensional spherically symmetric black hole surrounded
by {\it black rings}. Various aspects of this blossomming subject of researches were treated in Refs.
\cite{obe08}-\cite{emp08} (see also references therein). 
\par
In our paper we shall examine the first law of thermodynamics for charged black p-brane
in $(n + p + 1)$-dimensional dilaton gravity with $(p + 2)$- form strength field. In what follows one finds
the {\it physical process} version of the first law of thermodynamics for charged black p-brane
and the so-called {\it equilibrium version} of this law.
\par
The {\it physical process} version of the first law of black object thermodynamics
can be established by changing a stationary black object 
by some infinitesimal physical process. For instance, it can be realized 
throwing matter into black object. In addition, we assume that the 
final state of black object settles down to a 
stationary one, then 
it will be possible to extract the changes of black object's parameters. This in turn,
enables one to gain information about the first law of its mechanics. 
The {\it physical process } version of the first 
law of black hole thermodynamics was widely studied in the context of Einstein and Einstein-Maxwell (EM)
theory in Refs.\cite{wal94,gao01} as well as
in Einstein-Maxwell axion-dilaton (EMAD) gravity being the low-energy limit of the heterotic
string theory in Ref.\cite{rog02}. While the case  Einstein gravity coupled to
$(n-2)$-gauge form field strength was treated in Ref.\cite{rog05}.
On the other hand, {\it black rings} case was examined in Ref.\cite{rog05br}.
\par
The {\it equilibrium state} version of the first law of black holes mechanics
constitutes the other attitude to the problem in question. 
It was studied in the seminal paper of
Bardeen, Carter and Hawking \cite{bar73}. This attempts are based on considering
the linear perturbations of a stationary electrovac black hole to another one.
Ref.\cite{sud92} was devoted to
arbitrary asymptotically flat perturbations of a stationary 
black hole, while the first law of black hole thermodynamics valid for
an arbitrary diffeomorphism invariant Lagrangian 
with metric and matter fields possessing stationary and axisymmetric
black hole solutions were obtained in Refs.\cite{wal93}-\cite{iye97}. The cases of higher 
curvature terms and higher derivative terms in the metric
were considered in
\cite{jac}, while the situation when the Lagrangian is an arbitrary function of metric,
Ricci tensor and a scalar field was elaborated in Ref.\cite{kog98}.
In Ref.\cite{gao03}
of a charged rotating black hole
where fields were not smooth through the event horizon was treated.
The first law of black hole thermodynamics was also intensively studied in the case of $n$-dimensional
black holes. The {\it equilibrium state} version was elaborated in Ref.\cite{equi1} 
under the assumption of spherical topology of black holes. Some of the works assume that four-dimensional black 
hole uniqueness theorem extends to higher dimensional case \cite{equi2}.
The ADM mass and Komar surface integrals for energy density, tension and angular momentum density
of a stationary p-brane were given in Ref.\cite{tow01}.
As far as the {\it black ring} first law of mechanics is concerned, 
the general form of this law was achieved in Ref.\cite{cop05}, using the notion of
bifurcate Killing horizons and considering dipole charges.
In $n$-dimensional gravity containing $(p + 1)$-form field strength and dilaton fields
the first law of black ring mechanics choosing an arbitrary cross
section of the event horizon to the future of the bifurcation surface was derived in Ref.\cite{rog05br1}.
In $n$-dimensional Einstein gravity with Chern-Simons term
the {\it physical process} version and the {\it equilibrium state} version of the first law
of black ring thermodynamics were derived in Ref\cite{rog07ch}, while the case of black Saturn was presented
in \cite{rog07}. 
On the other hand, by means of the covariant cohomological methods to the conserved charges
for $p$-form gauge fields coupled to gravity the first law of thermodynamics was found in Ref.\cite{com07}.
\par
The paper is organized as follows. We devoted Sec.II to the {\it physical process} version of
charged p-brane first law of thermodynamics. Sec.III studies the {\it equilibrium state} version of the first law
of thermodynamics choosing an arbitrary cross section of p-brane event horizon to the future
of their bifurcation surfaces. Such attitude enables one to take into account fields which are
not necessarily smooth through the event horizon of charged p-brane under consideration.
Sec.IV concludes our investigations.

\section{Physical process version of the first law of black hole mechanics}
We begin with the Lagrangian 
describing $(n + p + 1)$-dimensional dilaton gravity with $(p + 2)$-form strength fields.
It is subject to the relation as follows:
\be
{\bf L } = {\bf \ep} \bigg(
{}^{(d)}R - {1 \over 2} \na_{\mu}\phi \na^{\mu} \phi - {1\over 2 (p + 2)!} e^{-{\alpha} \phi}
F_{\mu_{1} \dots \mu_{p+2}}~F ^{\mu_{1} \dots \mu_{p+2}}
\bigg),
\label{lag}
\ee
where $ {\bf \ep}$ is the volume element of dimension $d = n + p + 1$,
$\phi$ is the dilaton field while
$F_{\mu_{1} \dots \mu_{p+2}} = (p + 2)! \na_{[ \mu_{1}} A_{{\mu_{2} \dots \mu_{p+1}]}} $ is $(p + 2)$-form field strength,
with a potential
$A_{\mu_{1} \dots \mu_{p + 1}}$. By $\alpha$ we have denoted an arbitrary dilaton coupling parameter.
One can check that equations of motion for the underlying theory are given by
\ben \label{m1}
G_{\mu \nu} - T_{\mu \nu}(F, \phi) &=& 0, \\
\na_{i_{1}}\bigg( e^{-{\alpha} \phi} F^{i_{1} \dots i_{p+2}} \bigg) &=& 0,\\ \label{m2}
\na_{\mu} \na^{\mu} \phi + {\alpha \over 2 (p + 2)!} e^{-{\alpha} \phi}
F_{\mu_{1} \dots \mu_{p+2}}~ F^{\mu_{1} \dots \mu_{p+2}} &=& 0.
\label{m3}
\een
On the other hand, the energy momentum tensor for $(p + 2)$-form field strength and dilatons has the form as
\ben
T_{\mu \nu}(F, \phi) &=&
{1 \over 2} \na_{\mu} \phi \na_{\nu} \phi - {1 \over 4} g_{\mu \nu} \na_{\alpha} \phi
\na^{\alpha} \phi \\ \nonumber
 &+& {1\over 2 (p + 2)!} e^{-{\alpha} \phi}
\bigg[ (p + 2) F_{\mu \nu_{2} \dots \nu_{p+1}}~ F_{\nu}{}{}^{\nu_{2} \dots \nu_{p+2}}
- {1 \over 2} g_{\mu \nu} F_{\mu_{1} \dots \mu_{p+2}}~ F^{\mu_{1} \dots \mu_{p+2}}
\bigg].
\een
In order to establish the {\it physical version} of the first law of p-brane thermodynamics 
we shall try to find the explicit expressions
for the variation of mass and angular momentum and the tension of the brane.
On evaluating
the variations of the Lagrangian (\ref{lag}) with respect to the adequate fields, we find that one finally obtain
\ben \label{dl}
\delta {\bf L} &=& {\bf \epsilon} \bigg(
G_{\mu \nu} - T_{\mu \nu}(F, \phi) \bigg)~ \delta g^{\mu \nu}
- {\bf \epsilon} \na_{j_{1}} \bigg( e^{-{\alpha} \phi} F^{j_{1} \dots j_{p+2}} \bigg) 
\delta A_{j_{2} \dots j_{p+2}}  \\ \nonumber
&+& {\bf \epsilon} \bigg( \na_{\mu}\na^{\mu} \phi + {\alpha \over 2 (p + 2)!} e^{-{\alpha} \phi}
F_{\mu_{1} \dots \mu_{p+2}}~ F^{\mu_{1} \dots \mu_{p+2}} \bigg)~\delta \phi
+ d {\bf \Theta}.
\een
Using the above formula (\ref{dl}) we get
the symplectic $(n + p)$-form
$\Theta_{j_{1} \dots j_{n+p}}[\psi_{\alpha}, \delta \psi_{\alpha}]$, which yields
\be
\Theta_{j_{1} \dots j_{n+p}}[\psi_{\alpha}, \delta \psi_{\alpha}] =
\ep_{m j_{1} \dots j_{n+p}} \bigg[
\omega^{m} - e^{-{\alpha} \phi} F^{m \nu_{1} \dots \nu_{p+1}}~\delta A_{\nu_{1} \dots \nu_{p+1}}
- \na^{m} \phi~ \delta \phi \bigg].
\label{jj}
\ee 
In relation (\ref{jj}) by $\psi_{\alpha}$ we denote fields in the considered theory
while $\delta \psi_{\alpha}$ is equal to their variations.
$\omega_{\mu}$ stands for the expression
\be
\omega_{\mu} = \na^{\alpha} \delta g_{\alpha \mu} - \na_{\mu} 
\delta g_{\beta}{}{}^{\beta}.
\ee
One can also remark that the adequate equations of motion can be read off relation
(\ref{dl}).
The standard procedure provide in Ref.\cite{gao01} 
enables one to identify variations of the fields with a general coordinate transformations 
induced by an arbitrary Killing vector field $\xi_{\alpha}$.
Next, one can find that  
the Noether $(n + p)$-form with respect to this above mentioned Killing vector, i.e., 
${\cal J}_{j_{1} \dots j_{n+p}} = \ep_{m j_{1} \dots j_{n+p}} {\cal J}^{m}
\big[\psi_{\alpha}, {\cal L}_{\xi} \psi_{\alpha}\big]$. The result of doing that is of the
following form:
\ben
{\cal J}_{j_{1} \dots j_{n+p}} &=& 
d \bigg( Q^{GR} + Q^B \bigg)_{j_{1} \dots j_{n+p}}
+ 2~ \ep_{m j_{1} \dots j_{n+p}} \bigg( G^{m}{}{}_{\eta} - T^{m}{}{}_{\eta}(F, \phi)
\bigg) \xi^{\eta} \\ \nonumber
&+& (p+1)~\ep_{m j_{1} \dots j_{n+p}}~ \xi^{d}~ A_{d \alpha_{3} \dots \alpha_{p+2}}~
\na_{\alpha_{2}} \bigg( e^{-{\alpha} \phi} F^{m \alpha_{2} \dots \alpha_{p+2}} \bigg),
\een
where $Q_{j_{1} \dots j_{n+p-1}}^{GR}$ yields
\be
Q_{j_{1} \dots j_{n+p-1}}^{GR} = - \ep_{j_{1} \dots j_{n+p-1} a b} \na^{a} \xi^{b},
\label{grn}
\ee
while $Q_{j_{1} \dots j_{n+p-1}}^{A}$ has the following form:
\be
Q_{j_{1} \dots j_{n+p-1}}^{A} = 
{{p + 1 \over (p + 2)!}}~ \ep_{m k j_{1} \dots j_{n+p-1}}
~\xi^{d}~A_{d \alpha_{3} \dots \alpha_{p+2}}~ e^{-{\alpha} \phi} F^{m k \alpha_{3} \dots \alpha_{p+2}}.
\label{an}
\ee
Having in mind that ${\cal J}[\xi] = dQ[\xi] + \xi^{\alpha} {\bf C}_{\alpha}$,
where ${\bf C}_{\alpha}$ is an $(n+p)$-form constructed from dynamical fields, i.e., from
$g_{\mu \nu}$, $(p + 2)$-form field $F^{j_{1} \dots j_{p+2}}$ and dilaton fields,
one may consequently identify the sum of
relations (\ref{grn}) and (\ref{an}) with the Noether charge for 
$(n + p +1)$-dimensional dilaton gravity theory with $(p + 2)$-form strength field.
${\bf C}_{\alpha}$ implies
\be
C_{k j_{1} \dots j_{n+p}} = 2~ \ep_{g j_{1} \dots j_{n+p}}
\bigg[ G_{k}{}{}^{g} - T_{k}{}{}^{g}(F, \phi) \bigg] +
(p + 1)~ \ep_{g j_{1} \dots j_{n+p}}~\na_{\alpha_{2}} \bigg( 
e^{-{\alpha} \phi} F^{g \alpha_{2} \dots \alpha_{p+2}} \bigg)
A_{k \alpha_{3} \dots \alpha_{p+2}}.
\ee
The source-free Eqs. of motion are provide by the requirement that
 ${\bf C}_{\alpha} = 0$. On the contrary,
when the source-free equations do not hold yield the following relations:
\ben
G_{\mu \nu} - T_{\mu \nu}(F, \phi) &=& T_{\mu \nu}(matter) , \\
\na_{\xi_{1}} \bigg( e^{-{\alpha} \phi} F^{\xi_{1} \dots \xi_{p+2}} \bigg)
 &=& j^{\xi_{2} \dots \xi_{p+2}}(matter).
\een
If one further assumes that $(g_{\mu \nu},~ A_{\alpha_{1} \dots \alpha_{p+1}},~ \phi)$ are solutions 
of source-free equations of motion and 
$(\delta g_{\mu \nu},~\delta A^{\alpha_{1} \dots \alpha_{p+1}},~\delta \phi)$
are the linearized perturbations satisfying Eqs. of motion with sources
$\delta T_{\mu \nu}(matter)$ and $ \delta j^{\mu_{1} \dots \mu_{p+1}}(matter)$, then 
one obtains the relation of the form as
\be
\delta  C_{k m_{1} \dots m_{n+p}} = 2~ \ep_{g m_{1} \dots m_{n+p}}
\bigg[ \delta T_{k}{}{}^{g}(matter) + (p + 1)~A_{k \alpha_{3} \dots \alpha_{p+2}}~ 
\delta j ^{g \alpha_{3} \dots \alpha_{p+2}}(matter) \bigg].
\ee
Because of the fact that the Killing vector field $\xi_{\alpha}$ describes also a symmetry of the background
matter field, one gets the formula for a conserved quantity connected with $\xi_{\alpha}$, namely
\ben \label{hh}
\delta H_{\xi} &=& - 2~ \int_{\Sigma}\ep_{m j_{1} \dots j_{n+p}} \bigg[
\delta T_{a}{}{}^{m}(matter) \xi^{a} + 
 (p+1)~\xi^{i}~A_{i \alpha_{3} \dots \alpha_{p+2}}~ 
\delta j ^{m \alpha_{3} \dots \alpha_{p+2}}(matter) \bigg]
 \\ \nonumber
&+& \int_{\p \Sigma}\bigg[
\delta Q(\xi) - \xi \cdot \Theta \bigg].
\een
Let us 
choose $\xi^{\alpha}$ to be an asymptotic time translation $t^{\alpha}$, then
one can conclude that
$M = H_{t}$ and finally obtain 
the variation of the ADM mass
\ben \label{mm}
\alpha~ \delta M &=& - 2 \int_{\Sigma} \ep_{m j_{1} \dots j_{n+p}} \bigg[
\delta T_{k}{}{}^{m}(matter)~ t^{k} + (p+1)~ t^{k}~ A_{k \alpha_{3} \dots \alpha_{p+2}}~ 
\delta j ^{m \alpha_{3} \dots \alpha_{p+2}}(matter) \bigg] \\ \nonumber
&+& \int_{\p \Sigma}\bigg[
\delta Q(t) - t \cdot \Theta \bigg],
\een
where $\alpha = {n-2 \over n-1}$.
Next, if we take the Killing vector fields $\varphi_{(i)}$ which are responsible
for the rotation in the adequate directions, we arrive at the relations for angular 
momenta
\ben \label{jjj}
\delta J_{(i)} &=& 2 \int_{\Sigma} \ep_{m j_{1} \dots j_{n+p}} \bigg[
\delta T_{a}{}{}^{m}(matter) \varphi_{(i)}^{a} + (p+1)~ \varphi_{(i)}^{a}~ A_{a \alpha_{3} \dots \alpha_{p+2}}~ 
\delta j ^{m \alpha_{3} \dots \alpha_{p+2}}(matter) \bigg] \\ \nonumber
&-& \int_{\p \Sigma}\bigg[
\delta Q({\varphi}_{(i)}) - {\varphi}_{(i)} \cdot \Theta \bigg].
\een
Moreover, p-brane spacetime may have more spatial Killing vector fields tangent to the brane.
Just, consequently with the above definition we introduce 
$p$ translational Killing vectors $l^{\alpha}_{(j)}$ which are connected
with the p-brane tension in the adequate direction. Thus, the change of the brane tension in the adequate direction yields
\ben \label{tt}
\delta \cT_{(i)} &=& - 2 \int_{\Sigma} \ep_{m j_{1} \dots j_{n+p}} \bigg[
\delta T_{k}{}{}^{m}(matter)~ l^{k}_{(i)} + (p+1)~ l^{k}_{(i)}~ A_{k \alpha_{3} \dots \alpha_{p+2}}~ 
\delta j ^{m \alpha_{3} \dots \alpha_{p+2}}(matter) \bigg] \\ \nonumber
&+& \int_{\p \Sigma}\bigg[
\delta Q(l_{(i)}) - l_{(i)} \cdot \Theta \bigg],
\een
Let us suppose further that a stationary p-brane solutions, we consider, is
regular on and outside event horizon. Moreover, let us assume that the p-brane
event horizon is a Killing  horizon, which implies that there exists a Killing
vector field $\chi^{\alpha}$ normal to it. 
The Killing vector field $\chi^{\mu}$ is of the form as
\be
\chi^{\mu} = t^{\mu} + \sum_{i} \Omega_{(i)} \varphi^{\mu (i)}.
\label{kkh}
\ee
The surface integrals in Eqs.(\ref{mm})-(\ref{tt}) we understand as a surface
integrals over p-brane event horizon and a bulk integral over the region bounded by the
considered horizon and transverse spatial infinity on a $n$-surface having vector $t^{\delta}$
as one of its $(p + 1)$ normals. 
One has in mind that we assume existence of Killing vector field responsible for stationarity $t^{\delta}$,
Killing vector fields $\varphi^{\mu (i)}$ which are connected with rotation in the adequate direction as well as 
Killing vector fields bounded to the translation in various directions. We consider both 
{\it homogeneous} p-brane when $l_{(i)}$ are the same in every direction and {\it inhomogeneous} brane 
when they differ in every directions. We would call these states of p-brane
{\it translationary invariant} (for brevity), having in mind what was written above.
All Killing vector fields in question are mutually commutating.
\par
Let us perturb the black p-brane by dropping in some matter and assume that in the process
of this action p-brane will be not destroyed and settle down to a stationary
and {\it translationary invariant} final state. Then, the next task will be to
find the changes of the black p-brane parameters.
Changes of the event horizon area of a p-brane will be computed by means of
$n$-dimensional Raychaudhuri equation.
In addition, we shall assume that
$\Sigma_{0}$ is an asymptotically flat
hypersurface which terminating on the p-brane event horizon.
Then, one takes into account 
the initial data on $\Sigma_{0}$
for a linearized perturbations of
$(\delta g_{\mu \nu},~ \delta A_{\alpha_{1} \dots \alpha_{p+1}}, \delta \phi)$
with $\delta T_{\mu \nu}(matter)$ and $\delta j^{\alpha_{1} \dots \alpha_{p+1}}(matter)$. 
We require that $\delta T_{\mu \nu}(matter)$ and $\delta j^{\alpha_{1} \dots \alpha_{p+1}}(matter)$
disappear
at infinity and the initial data for 
$(\delta g_{\mu \nu},~ \delta A_{\alpha_{1} \dots \alpha_{p+1}}, \delta \phi)$
vanish in the vicinity of black p-brane event horizon $\cal H$ on 
the adequate
hypersurface $\Sigma_{0}$.
The above conditions provide
that for the initial time
$\Sigma_{0 }$, considered black p-brane is unperturbed. On its own, it causes that
the perturbations vanish near the internal boundary $\p \Sigma_{0}$.
From relations (\ref{mm}) and (\ref{tt}) one gets the following is fulfilled:
\ben \label{ppp}
\alpha~ \delta M &-&  \sum_{i} \Omega_{(i)} \delta J^{(i)} - 
\sum_{i} \delta \cT_{(i)} = \\ \nonumber
&-& 2 \int_{\Sigma_{0}} \ep_{m j_{1} \dots j_{n+p}} \bigg[
\delta T_{f}{}{}^{m}(matter)~ \chi^{f} + (p + 1)~ \chi^{k}~ A_{k \alpha_{2} \dots \alpha_{p+1}}~ 
\delta j ^{m \alpha_{2} \dots \alpha_{p+1}}(matter) \bigg] \\ \nonumber
&-& 2 \sum_{i}
\int_{\Sigma_{0}} \ep_{m j_{1} \dots j_{n+p}} \bigg[
\delta T_{k}{}{}^{m}(matter)~ l^{k}_{(i)} + (p+1)~ l^{k}_{(i)}~ A_{k \alpha_{3} \dots \alpha_{p+2}}~ 
\delta j ^{m \alpha_{3} \dots \alpha_{p+2}}(matter) \bigg] \\ \nonumber
&=& \int_{\cH} \gamma ^{\alpha}~k_{\alpha}~\bep_{j_{1} \dots j_{n-1}}~v(l),
\een
where $\bep_{j_{1} \dots j_{n-1}} = n^{\delta}~\ep_{\delta j_{1} \dots j_{n-1}}$ and
$v(l) = \int \ep_{j_{1} \dots j_{p}}$. By
$n^{\delta}$ we denoted the future directed unit normal to the hypersurface $\Sigma_{0}$.
$k_{\alpha}$ is a tangent vector to the affinely parametrized null geodesics generators of 
p-brane  event horizon. Further, we assume that all of the matter falls into the considered 
black p-brane. We have also in mind that the current $\gamma^{\alpha}$ is conserved.
Due to the above facts we replace in relation (\ref{ppp}) vector $n^{\beta}$ by the vector $k^{\beta}$ defined above.
\par
We shall assume that the field strength $F_{\alpha_{1} \dots \alpha_{p+2}}$ is invariant under
symmetries generated by adequate Killing vector fields. Namely, the adequate Lie derivatives
of gauge field $A_{\alpha_{1} \dots \alpha_{p+1}}$  are equal to zero. One gets the following:
\be
\cL_{\chi} A_{\alpha_{1} \dots \alpha_{p+1}} = 0, \qquad 
\cL_{l} A_{\alpha_{1} \dots \alpha_{p+1}} = 0.
\ee
The same relations are satisfied by dilaton field
\be
\cL_{\chi} \phi = 0, \qquad \cL_{l} \phi = 0.
\ee
It can be checked by the direct calculations that for $\xi^{\alpha}$
generating symmetries of the considered background 
the following
relation takes place:
\ben \label{cf}
(p + 1)!~\cL_{\xi} A_{\alpha_{1} \dots \alpha_{p+1}}~\delta j^{\alpha_{1} \dots \alpha_{p+1}}
&-& \xi^{d}~F_{d \alpha_{2} \dots \alpha_{p+2}}~\delta j^{\alpha_{2} \dots \alpha_{p+2}} \\ \nonumber
&=& (p + 1)~ (p + 1)! \na_{\alpha_{2}} \bigg(
\xi^{d}~A_{d \alpha_{3} \dots \alpha_{p+2}}
\bigg)~\delta j^{\alpha_{2} \dots \alpha_{p+2}}.
\een
Eq.(\ref{cf}) will be useful in calculations of the integral over black p-brane event horizon.
In stationary background expansion $\theta$ and shear $\sigma_{ij}$ will vanish.
Using higher dimensional Raychaudhuri equation of the form
\be
{d \theta \over d \lambda} = - {\theta^{2} \over (n + p - 1)} - \sigma_{ij} \sigma^{ij}
- R_{\mu \nu} \xi^{\mu} \xi^{\nu},
\label{ray}
\ee
where $\lambda$ denotes the affine parameter corresponding to vector $k_{\alpha}$, one
concludes that
$R_{\alpha \beta} k^{\alpha} k^{\beta} \mid_{\cH} = 0$.
Due to this fact we get relation of the form as
\be
{1 \over 2}~k^{\mu} \na_{\mu} \phi~ k^{\nu}~ \na_{\nu} \phi +
{1 \over 2 (p + 1)!} e^{- \alpha \phi}~
F_{\mu \mu_{2} \dots \mu_{p+2}}~ F_{\nu}{}{}^{ \mu_{2} \dots \mu_{p+2}}~ k^{\mu}~ k^{\nu} \mid_{\cH} = 0.
\ee
Using the fact that $\cL_{k} \phi = 0$,
it is easily seen that,
$F_{\alpha}{}{}^{ \mu_{2} \dots \mu_{p+2}} k^{\alpha} = 0$. Because of the fact that
$F_{\alpha \mu_{2} \dots \mu_{p+2}} k^{\alpha} k^{\mu_{2}} = 0$, by asymmetry of $F_{ \mu_{1} \dots \mu_{p+2}}$
it turned out that
$F_{\alpha \mu_{2} \dots \mu_{p+1}} k^{\alpha} \sim k_{\mu_{2}} \dots k_{\mu_{p+1}}$.
It implies that the
pull-back of $F_{\alpha}{}{}^{ \mu_{2} \dots \mu_{p+2}} k^{\alpha}$ to the p-brane event horizon is equal to zero.
In turn, it reveals the fact that
$\chi^{k}~F_{k \alpha_{2} \dots \alpha_{p+2}}$ is a closed
$(p+1)$-form on the p-brane event horizon. 
\par
The same considerations as above may be applied to the integrals concerning with p-brane tension.\\
Now, we proceed to the surface terms. It follows that the adequate surface terms will have form of 
$\Phi~\delta Q$, where $\Phi$ is the constant sum relating to the harmonic parts
of $\xi^{d}~F_{d \alpha_{2} \dots \alpha_{p+2}}$ and $\delta Q$ is the variation of local charges.
These allow one to write down the following:
\be
\alpha~ \delta M -  \sum_{i} \Omega_{(i)}~ \delta J^{(i)} - \sum_{i} \delta \cT_{(i)}
+ \Phi~\delta Q =
4 \int_{\cH}
\delta T_{\mu}{}{}^{\nu} \xi^{\mu}~ k_{\nu}~v(l),
\label{rh}
\ee
where $\Phi~\delta Q = \Phi_{(\chi)}~\delta Q_{(\chi)} + \sum_{i}\Phi_{l_{(i)}}~\delta Q_{l_{(i)}}$
is the sum of the potentials and local charges connected with the adequate Killing fields.
Our next task is to find
the right-hand side of Eq.(\ref{rh}). It can be elaborated by
the same procedure as described in Refs.\cite{gao01,rog02,rog05}. Namely,
considering $(n+ p +1)$-dimensional Raychaudhuri Eq. and 
using the fact that
the null generators of the event horizon of the perturbed black p-brane coincide with
the null generators of the unperturbed black p-brane, leads to the relation of the form
\be
\kappa~ ~\delta \cA_{eff} = \int_{\cal H} \delta
T^{\mu}{}{}_{\nu}(matter)~ \xi^{\nu}~ k_{\mu}~v(l),
\label{kap}
\ee
where $\kappa$ is the surface gravity of black p-brane while
$\delta \cA_{eff} = v(l)~\delta \cA$, is just an $(n - 1)$-dimensional effective area of the event horizon
of the considered p-brane. The same reasoning enables us to find the same expression when on the right-hand side
of Eq.(\ref{kap}) are $l_{(i)}$ instead of $\xi^{\nu}$.\\

In the light of what has been shown above we arrive at the
{\it physical process} version of the first law of black p-brane 
mechanics in Einstein $(n + p + 1)$-dimensional gravity with
additional
$(p + 2)$-form field strength and dilaton fields. It is provided by
\be
\alpha~ \delta M - \sum_{i} \Omega_{(i)} \delta J^{(i)} - \sum_{i} \delta \cT_{(i)}
+ \Phi~\delta Q
 = 4 \kappa ~\delta \cA_{eff}.
\ee
We finally remark that in the sense of Ref.\cite{gao01} the proof of {\it physical process} version of the first law of
thermodynamics for $(n + p + 1)$-dimensional black p-brane also provides support for cosmic censorship.

\section{Equilibrium State Version of the First Law of p-brane Mechanics} 
In this section we shall look for the {\it equilibrium state} version of the first law of
charged black p-brane thermodynamics. In Ref.\cite{wal00} it was shown that in the spacetime
with asymptotic conditions at infinity and possessing Killing vector field $\xi_{\mu (i)}$
which generates asymptotical symmetry it will be possible to define the {\it conserved}
quantity $H_{\xi (i)}$, which is given by the relation
\be
\delta H_{\xi (i)} = \int_{\infty} \bigg( \bdel Q(\xi_{(i)}) - \xi_{(i)} \Theta \bigg).
\label{qua}
\ee
$\bdel$ denotes variation which has no effect on $\xi_{\alpha}$ since the Killing
vector field in question is treated as a fixed background and it should not to be varied in the
above expression (\ref{qua}).
\par
In our considerations we shall take into account $(n + p + 1)$-dimensional spacetime with charge p-brane.
As was mentioned in the preceding section we have to do with the Killing vector field
$\chi^{\mu}$ which is normal to the p-brane event horizon and translational Killing vectors $l^{\mu}_{(i)}$
in addition to the aforementioned ones. Moreover, we assume that all are mutually commutating.\\
In what follows we choose an arbitrary cross section of the considered p-brane event horizon
to the future of the bifurcation surface. In Ref.\cite{gao03} it was revealed that
such attitude enabled one to 
treat fields which were not necessarily smooth through the event horizon of the black object.
The only requirement is that the pull-back
of these fields in the future of the bifurcation surface be smooth. 
\par
To derive {\it equilibrium state} version of the first law of charged p-brane mechanics
let us consider asymptotically hypersurfaces 
$\Sigma$ ending on the part of the p-brane event horizons $\cal H$ 
to the future of the bifurcation surfaces. 
The inner boundary $S_{\cal H} $
of the hypersurface $\Sigma$ will be the cross sections of the black p-brane
event horizon. 
Next, we shall compare variations between two neighbouring states of the p-brane.
One should recall \cite{bar73} that
there is a freedom which points can be chosen to correspond when one compares two
slightly different solutions.
In our consideration we choose the freedom of the generalized coordinate transformation
and put $S_{\cal H}$ the same of the two solutions 
Moreover, one takes into account the case when the null vector remains normal to $S_{\cal H}$.
The stationarity, axisymmetricity 
and translantionarity of the considered solutions will be conserved, which provides in turn that
$\delta t^{\mu}$, $\delta \varphi^{\mu (i)}$ , and $\delta l^{\mu}$ will be equal to zero.\\
On the other hand, the variation of the Killing vector field $\chi_{\mu}$ normal to the charged p-brane event
horizon will be given by the following:
\be
\delta \xi^{\mu} = \sum_{i} \delta \Omega_{(i)} \varphi^{\mu (i)}.
\ee
Let us suppose that 
$(g_{\mu \nu}, A_{\alpha_{1} \dots \alpha_{p+1}}, \phi)$ are solutions 
of the equations of motion and their variations
$(\delta g_{\mu \nu},~\delta A^{\alpha_{1} \dots \alpha_{p+1}},~\delta \phi)$ constitute
their linearized perturbations also fulfill Eqs. of motion. 
One requires also that the pull-back
of the potential $A_{\alpha_{1} \dots \alpha_{p+1}}$ to the future of the bifurcation surface be smooth, but not 
necessarily smooth on it \cite{gao03}. We require further that $A_{\alpha_{1} \dots \alpha_{p+1}}$
and its variation $\delta A_{\alpha_{1} \dots \alpha_{p+1}}$ vanish sufficiently rapid at infinity.
Consequently, for charged black p-brane one obtains
\be
\alpha 
\delta M
-  \sum_{i} \Omega_{(i)} \delta J^{(i)} - \sum_{i} \delta \cT_{(i)}
=
\int \bigg( \bdel Q(\chi) - \chi~ \Theta \bigg)
- \int \bigg( \bdel Q(l_{(i)}) - l_{(i)}~ \Theta \bigg).
\label{tens}
\ee
To begin with we shall find the integral over symplectic $(n + p)$-form connected with
dilaton field. In the case under consideration the volume element
has the  form
\be
\ep_{\mu a j_{1} \dots j_{n+p-1}} = \chi_{\mu} \wedge N_{a} \wedge \ep_{j_{1} \dots j_{n-1}}
\wedge \ep_{j_{1} \dots j_{p}},
\ee
where vector $N_{\beta}$ is the {\it ingoing} future directed null normal to the
p-brane event horizon $S_{\cal H}$. It is normalized as follows:
\be
N^{\mu}~\chi_{\mu} = -1.
\ee
Just, we arrive at the relation of the form
\be
\int \chi^{j_{1}}~ \Theta_{j_{1} \dots j_{n+p}}^{\phi} =
\int_{S_{\cal H}}~v(l)~ \ep_{j_{1} \dots j_{n-1}}~ N_{\alpha } \chi^{\alpha}~ \chi_{\mu } \na^{\mu} \phi~
\delta \phi = 0,
\ee
where we used the fact that $\cL_{\chi} \phi = 0$.\\
The arguments presented in the
preceding section can be applied now. It leads to the following:
\be
\int Q_{j_{1} \dots j_{n+p-1}}^{A}(\chi) 
= \Phi_{(\chi)}~Q_{(\chi)}. 
\ee
Our next task will be to find the variation $\bdel$ of $Q_{j_{1} \dots j_{n+p-1}}^{A}(\chi)$.
Then, one obtains
\ben \label{bar}
\bdel \int Q_{j_{1} \dots j_{n+p-1}}^{A}(\chi) &=& \noindent \\
 &=& \delta (\Phi_{(\chi)}~Q_{(\chi)}) 
- {(p + 1)~v(l) \over (p + 2)!} \int_{S_{\cal H}} \sum_{i} \delta \Omega_{(i)} \varphi^{\mu (i)}
A_{\mu \alpha_{3} \dots \alpha_{p+2}} 
\ep_{m k  j_{1} \dots j_{n-2}}~ e^{-{\alpha} \phi}~ 
F^{m k \alpha_{3} \dots \alpha_{p+2}} 
\een
As a direct consequence of relation (\ref{bar}) we arrive at the expression which can be written as
\ben \label{pt}
\delta \Phi_{(\chi)}~Q_{(\chi)} 
&=& {(p + 1)~v(l) \over (p + 2)!}\int_{S_{\cal H}} \sum_{i} \delta \Omega_{(i)}~\varphi^{\mu (i)}~
A_{\mu \alpha_{3} \dots \alpha_{p+2}}~ 
\ep_{m j j_{1} \dots j_{n-2}}~ e^{-{\alpha} \phi}~ 
F^{m j \alpha_{3} \dots \alpha_{p+2}} \\ \nonumber
&+& {(p + 1)~v(l) \over (p + 2)!}
\int_{S_{\cal H} } \chi^{d}~\delta A_{d \alpha_{3} \dots \alpha_{p+2}}
~N_{m}~\chi_{j }~e^{-{\alpha} \phi}~ 
F^{m j \alpha_{3} \dots \alpha_{p+2}}.
\een
Using the fact that on the event horizon of black p-brane  
$F_{\mu \mu_{2} \dots \mu_{p+2}} \chi^{\mu} \sim \chi_{\mu_{2}} \dots \chi_{\mu_{p+1}}$
and expressing $\ep_{\mu a j_{1} \dots j_{n-2}}$ in the same form
as in the above case, one gets the following:
\be 
\int \chi^{j_{1}}~\Theta_{j_{1} \dots j_{n+p}}^{A} 
= {(p + 1)~v(l) \over (p+2)!}
\int_{S_{\cal H}} \ep_{j_{1} \dots j_{n-2}}~ e^{-{\alpha} \phi}~ 
\chi_{k}~ F^{j k \nu_{3} \dots \nu_{p+2}}~N_{j}~
\chi^{\nu_{2}}~ \delta A_{\nu_{2} \dots \nu_{p+2}} 
 \label{bb1}
\ee
Having in mind Eqs.(\ref{pt}) and (\ref{bb1}) 
one can conclude that 
\be
\bdel \int Q_{j_{1} \dots j_{n+p-1}}^{A} (\chi)
- \chi^{j_{1}}~\Theta_{j_{1} \dots j_{n+p}}^{A} 
= 
\Phi_{(\chi)}~\delta Q_{(\chi)}. 
\label{char}
\ee
Now, let us turn our attention to the contribution bounded with gravitational field. Namely,
for p-brane one obtains
\be
\int Q_{j_{1} \dots j_{n+p-1}}^{GR} (\chi) = 2 \kappa~ \cA_{eff},
\ee
where $\cA_{eff} = v(l)~\int_{S_{\cal H}} \ep_{j_{1} \dots j_{n-1}}$ 
is the area of the p-brane event horizon.\\
Then, it implies  
\be
\bdel \int Q_{j_{1} \dots j_{n+p-1}}^{GR} (\chi) 
=
2 \delta \bigg( \kappa~ \cA_{eff} \bigg)
+ 2 \sum_{i} \delta \Omega_{(i)}~ J^{(i)}
\ee
where we have denoted by
$J^{(i)}= {1 \over 2}\int_{S_{\cal H}}~v(l)~\ep_{j_{1} \dots j_{n-2} a b} \na^{a} \varphi^{(i) b}$
the angular momentum connected with the Killing vector
fields responsible for the rotations in the adequate directions.
Following the calculations presented in Ref.\cite{bar73} it could be found that
the following integral is satisfied:
\be
\int \chi^{j_{1}}~ \Theta_{j_{1} \dots j_{n+p-1}}^{GR} (\chi) 
=
2~\cA_{eff}~ \delta \kappa + 2 \sum_{i} \delta \Omega_{(i)}~ J^{(i)}.
\ee                 
The above relation yields the conclusion that
\be
\bdel \int Q_{j_{1} \dots j_{n+p-1}}^{GR} (\chi)
- \chi^{j_{1}}~\Theta_{j_{1} \dots j_{n+p}}^{GR} 
= 
2 \kappa~\delta \cA_{eff}                    
+ 2 \sum_{a} \kappa~\delta \cA_{eff}.
\label{arr}
\ee
The entirely analogous considerations can be applied to the second part of the right-hand side of Eq.(\ref{tens})
related to the brane tension and Killing vector fields $l_{(i)}$.
\par
Thus, the 
direct consequence of relations (\ref{char}) and (\ref{arr}) 
and the analogous for brane tension integrals
provides
the first law of charged black p-brane mechanics in Einstein
$(n + p +1)$-dimensional gravity with additional $(p+2)$-form field strength and dilaton
fields. The first law of mechanics for the considered black objects can be written in the form as    
\be
\alpha~ 
\delta M
-  \sum_{i} \Omega_{(i)} \delta J^{(i)} 
+
\Phi~\delta Q - \sum_{i} \delta \cT_{(i)}
 = 4 \kappa ~\delta \cA_{eff}, 
\ee
where $\Phi$ and $\delta Q$ are the adequate sums of constant potentials on $S_{\cal H}$ and sum of local charges.

\section{Conclusions}
In our paper we studied the first the first law of 
charged black p-brane thermodynamics in $(n + p +1)$-dimensional dilaton gravity with
$(p+2)$-form field strength. We assumed stationarity and axisymmetricity of the
considered p-brane. Moreover, we supposed that there were $p$ translation Killing vectors in addition
to $t^{\mu}$ Killing vector field and $\varphi_{(i)}^{\mu}$ Killing vectors responsible for the rotations 
in the adequate directions. All these Killing vectors commute mutually.
We looked for both {\it physical process} version and {\it equilibrium state} version
of the first law of charged p-brane thermodynamics.
\par
Considering {\it physical process} version of the first law of p-brane dynamics 
we change infinitesimally the p-brane under consideration by throwing matter into it.
Assuming that this process will not destroy black object in question we find changes of the
ADM mass, angular momentum, tension and effective area of the event horizon of the p-brane.
As far as {\it equilibrium state} version of first law of p-brane thermodynamics is concerned
we chose arbitrary cross sections of p-brane event horizons to the future
of bifurcation surfaces, contrary to the previous derivations which are bounded to the considerations
of bifurcation surfaces as the boundaries of hypersurfaces extending to spatial infinity.
It turn, such 
attitude enables one to treat fields which are not necessary smooth through each event horizon of the adequate
black object.
\par
As was shown in Ref.\cite{tra04}
the modification of the derivation of the first law of black p-brane
thermodynamics using the ADM-formalism was fruitful. It will be not
amiss to use this idea in higher dimensional p-brane spacetime. Perhaps
the reasoning presented in \cite{rog}
will be useful. We hope to return to the problem in question elsewhere.

\begin{acknowledgments}
This work was partially financed by the Polish budget funds in 2009 year as
the research project.
\end{acknowledgments}




\end{document}